\documentclass[draftcls,12pt,onecolumn,oneside]{IEEEtran}
\usepackage{mathrsfs}
\usepackage{algorithm}
\usepackage{algorithmic}
\usepackage{amsthm}
\usepackage{acronym}
\usepackage{mathrsfs}
\usepackage{ifthen}
\usepackage{textcomp}
\usepackage{float}
\usepackage{subfigure}
\usepackage{color}
\usepackage{graphicx, epstopdf}
\usepackage{amssymb, amsmath}
\usepackage{commath, cite}
\usepackage{setspace}
\usepackage{url,times,amsmath,color,amssymb,graphicx,epsfig,psfig,cite,geometry,psfrag, subfigure,dsfont,booktabs, multirow}

\begin{document}
\title{Coexistence of Wi-Fi and Heterogeneous Small Cell Networks Sharing Unlicensed Spectrum}

\author{Haijun Zhang, Xiaoli Chu, Weisi Guo and Siyi Wang
\thanks{Haijun Zhang is with College of Information Science and Technology, Beijing University of Chemical Technology, Beijing 100029, P. R. China, and is also with the Department of Electrical and Computer Engineering, the University of British Columbia, Vancouver, BC, V6T 1Z4, Canada  (Email: dr.haijun.zhang@ieee.org).
Xiaoli Chu is with Department of Electronic and Electrical Engineering, the University of Sheffield, Sheffield S1 3JD, UK (Email: x.chu@sheffield.ac.uk).
Weisi Guo is with School of Engineering, University of Warwick, CV4 7AL, UK (Email: weisi.guo@warwick.ac.uk).
Siyi Wang is with the Department of Electrical and Electronic Engineering, Xi'an Jiaotong-Liverpool University, China; and the Institute for Telecommunications Research, University of South Australia, Australia. (Email: siyi.wang@xjtlu.edu.cn).

}} \maketitle

\begin{abstract}
As two major players in terrestrial wireless communications, Wi-Fi systems and cellular networks have different origins and have largely evolved separately. Motivated by the exponentially increasing wireless data demand, cellular networks are evolving towards a heterogeneous and small cell network architecture, wherein small cells are expected to provide very high capacity. However, due to the limited licensed spectrum for cellular networks, any effort to achieve capacity growth through network densification will face the challenge of severe inter-cell interference. In view of this, recent standardization developments have started to consider the opportunities for cellular networks to use the unlicensed spectrum bands, including the 2.4 GHz and 5 GHz bands that are currently used by Wi-Fi, Zigbee and some other communication systems. In this article, we look into the coexistence of Wi-Fi and 4G cellular networks sharing the unlicensed spectrum. We introduce a network architecture where small cells use the same unlicensed spectrum that Wi-Fi systems operate in without affecting the performance of Wi-Fi systems. We present an almost blank subframe (ABS) scheme without priority to mitigate the co-channel interference from small cells to Wi-Fi systems, and propose an interference avoidance scheme based on small cells estimating the density of nearby Wi-Fi access points to facilitate their coexistence while sharing the same unlicensed spectrum. Simulation results show that the proposed network architecture and interference avoidance schemes can significantly increase the capacity of 4G heterogeneous cellular networks while maintaining the service quality of Wi-Fi systems.
\end{abstract}


\section{Introduction}
In recent years, the mobile data usage has grown by 70--200\% per annum. More worryingly, the bursty nature of wireless data traffic makes traditional network planning for capacity obsolete. Amongst both operators and vendors alike, small cells (e.g., picocells, femtocells and relay nodes) have been considered as a promising solution to improve local capacity in traffic hotspots, thus relieving the burden on overloaded macrocells.  A lot of research and development efforts have been made to efficiently offload excess traffic from macrocells to small cells, especially in indoor environments \cite{HwangIEEEmag2013}.

Due to the scarcity of licensed spectrum for cellular networks, small cells are expected to share the same spectrum with macrocells even when they are deployed within the coverage area of a
macrocell \cite{ChuCUP2013}. A frequency reuse factor of 1 in 3G HSPA+ and 4G LTE/LTE-A systems has proven
to yield high gains in network capacity. If without notable amounts of extra spectrum made
available for mobile communications, future cellular networks will unsurprisingly continue to
explore aggressive frequency reuse methods. Accordingly, the envisaged large-scale deployment
of small cells is likely to be hampered by the potentially severe co-channel interference between small cells and the umbrella macrocell and between neighboring small cells in dense deployment.

In view of this, the wireless industry is examining the efficient utilization of all possible spectrum resources including unlicensed spectrum bands to offer ubiquitous and seamless access to mobile users \cite{HuaweiUnlicensed131723}. The unlicensed 2.4 GHz and 5 GHz bands that Wi-Fi systems operate in have been considered as important candidates to provide extra spectrum resources for cellular networks. The initially targeted 5 GHz unlicensed band has potentially up to 500 MHz of spectrum available. In USA, Korea and China, deploying LTE-A in unlicensed spectrum does not require changes to the existing LTE-A standards (e.g., 3GPP Rel-10). In most other countries, the regulatory requirements of `Listen Before Talk' in unlicensed spectrum mandate standard modifications (e.g., candidates for 3GPP Rel-13).

Nowadays, most mobile devices such as smartphones and tablets support Wi-Fi connectivity, while the proliferation of Wi-Fi access points continues. The Wi-Fi access point density in developed urban areas has reached over 1000 per square km. Widely deployed Wi-Fi systems are playing an increasingly more important role in offloading data traffic from the heavily loaded cellular network, especially in indoor traffic hotspots and in poor cellular coverage areas. Very recently, the FCC voted to make 100 MHz of spectrum in the 5 GHz band available for unlicensed Wi-Fi use, giving carriers and operators more opportunities to push data traffic to Wi-Fi. Wi-Fi access points have even been regarded as a distinct tier of small cells in heterogenous cellular networks. However, since Wi-Fi systems are wireless local area networks (WLANs) based on the IEEE 802.11 standards, they have usually been designed and deployed independently from the cellular networks. Now that the wireless industry is seeking to explore the unlicensed spectrum currently used by Wi-Fi systems for LTE/LTE-A and future cellular networks' usage as well, the coexistence and interworking of Wi-Fi and heterogeneous cellular networks become an area requiring extensive research and investments. The joint deployment of Wi-Fi and cellular networks in the unlicensed spectrum can increase the overall capacity of a heterogeneous network, provided that the mutual interference between Wi-Fi and cellular systems is properly managed so that both can harmoniously coexist.

Benefits promised by the coexistence of Wi-Fi and cellular networks in unlicensed spectrum have started to attract interest from the research community. In \cite{Hajmohammad13ICC}, the authors proposed a quality of service (QoS) based strategy to split the unlicensed spectrum between Wi-Fi and femtocell networks. Although the unlicensed spectrum splitting scheme considers fairness between Wi-Fi access points and femtocells, the split use of the spectrum between two systems prohibits a high cross-network throughput.  In \cite{Charitos13ICC}, the authors investigated the deployment of a heterogeneous vehicular wireless network consisting of IEEE 802.11b/g and IEEE 802.11e inside a tunnel for surveillance applications, and specifically evaluated the handover performance of the hybrid Wi-Fi/WiMAX vehicular network in an emergency situation. In \cite{Almeida13ICC}, time-domain resource partitioning based on the use of almost blank subframes (ABSs) was proposed for LTE networks to share the unlicensed spectrum with Wi-Fi systems. Qualcomm has recently proposed to deploy LTE-A in the unlicensed 5 GHz band currently used mostly by Wi-Fi. The main idea is to deploy LTE-A as supplemental downlink (SDL) in the 5725-5850 MHz band in USA, with the primary cell always operating in the licensed band. Verizon and Ericsson are also exploring similar ideas. Huawei and CMCC have investigated the availability, commonality and feasibility of integrating the unlicensed spectrum to International Mobile Telecommunications-Advanced (IMT-A) cellular networks \cite{HuaweiUnlicensed131723}. LTE-Unlicensed (LTE-U) was first proposed by Qualcomm and Ericsson as a technology to run LTE in unlicensed spectrum in congested areas. Since February 2014, NTT DoCoMo and Huawei have been researching LTE-U, which they refer to as Licensed-Assisted Access using LTE (LAA-LTE). They have demonstrated on pre-commercial multi-cell networks that LAA-LTE achieves better coverage and capacity in the 5 GHz unlicensed spectrum than Wi-Fi alone. However, there are still concerns that LTE-U may completely take over the Wi-Fi bands in dense deployments.

It is worth noting that technical issues related to the coexistence of Wi-Fi and heterogeneous cellular networks in unlicensed spectrum, such as efficient spectrum sharing and interference mitigation, have not been sufficiently addressed. In \cite{Dimatteo11MASS}, the authors proposed  an integrated architecture exploiting the opportunistic networking paradigm to migrate data traffic from cellular networks to metropolitan Wi-Fi access points. In \cite{BennisIEEEmag2013}, Bennis \textit{et al.} introduced the basic building blocks of cross-system learning and provided preliminary performance evaluation in an LTE simulator overlaid with Wi-Fi hotspots. For the unlicensed spectrum sharing deployment of Wi-Fi and LTE-A systems, the co-channel interference between Wi-Fi tier and LTE-A tier can be mitigated by using ABSs, in which the interfering tier is not allowed to transmit data, and the victim tier can thus get a chance to schedule transmissions in the ABSs with reduced cross-tier interference~\cite{Liang12}. Moreover, it has been shown that by estimating the number of co-channel transmitters and knowing the deployment density of network nodes in a region, the average channel quality at any point in a coverage area can be inferred \cite{Guo14ICT}.

In this article, we present a network architecture to support the coexistence of Wi-Fi and heterogeneous cellular networks sharing the unlicensed spectrum. Based on the network architecture, we first provide an in-depth review of the ABS mechanism used for mitigating the co-channel interference from small cells to Wi-Fi systems and present a spectrum-sensing based fair ABS scheme without priority. We then propose an interference avoidance scheme based on small cells estimating the density of nearby Wi-Fi transmissions to facilitate the unlicensed spectrum sharing between small cells and Wi-Fi access points. Simulation results are provided to evaluate the performance of the proposed network architecture and interference avoidance scheme in facilitating coexistence of Wi-Fi and 4G heterogeneous cellular networks in unlicensed spectrum.
\begin{figure}[h]
        \centering
        \includegraphics*[width=15cm]{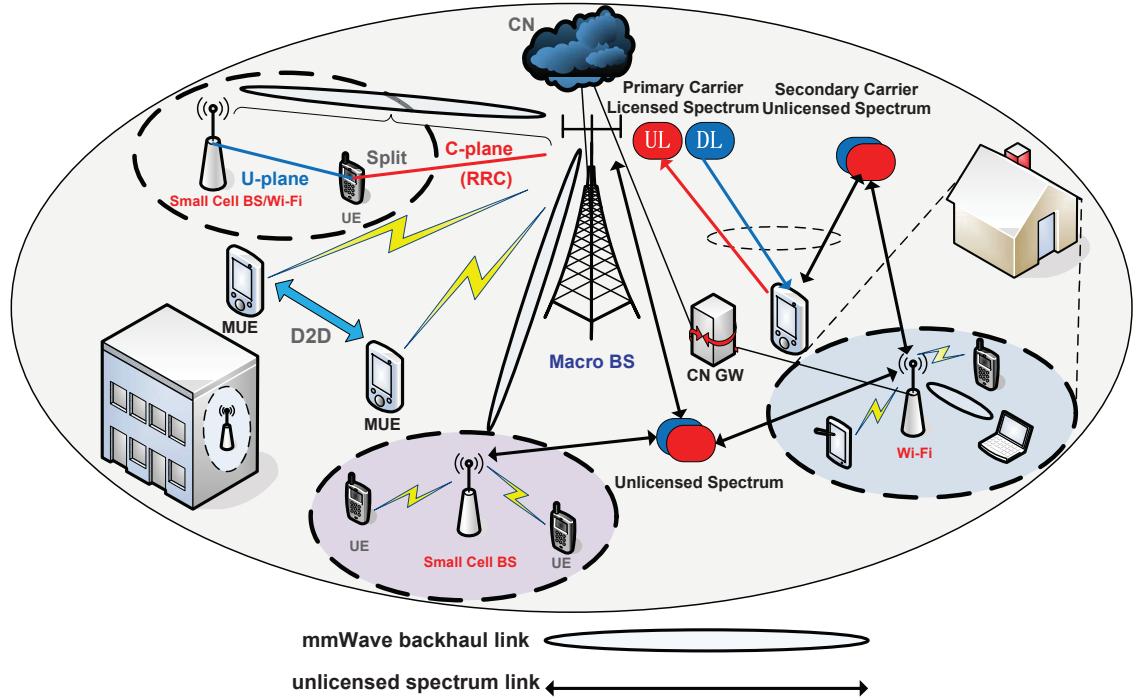}
        \caption{Network Architecture of LTE-A and Wi-Fi Coexistence.}
        \label{fig:1}
\end{figure}

\section{Network Architecture for Wi-Fi and Cellular Coexistence}

\subsection{Heterogeneous Network Architecture}

Fig. 1 shows the network architecture where several Wi-Fi access points and small cells coexist in the coverage area of a macrocell. The macrocell, small cells, and Wi-Fi access points share the same unlicensed spectrum for providing radio access to users, millimeter-wave radio is used for small-cell backhaul links, and device-to-device (D2D) communications are supported based on Wi-Fi Direct or LTE Direct.  As shown in Fig. 1, the control plane (C-plane) and user plane (U-plane) are split on the radio links associated with small cells. Specifically, the C-plane of  user equipments (UEs) associated with a  small cell is provided by the macro evolved Node B (eNB) in a low frequency band, while the U-plane of UEs associated with a small cell is provided by their serving small cell in a high frequency band. For UEs associated with the macrocell, both the C-plane and U-plane of their radio links  are provided by the serving macrocell. Since the C-plane of small-cell UEs is managed by the macrocell, the radio resource control (RRC) signallings of small-cell UEs are transmitted from the  macrocell, and the handover signalling overhead between small cells and the macrocell can be much reduced \cite{TrendsIEEEmag13}. Regarding Wi-Fi access points as a type of small cells, the C-plane and U-plane split can be applied in a Wi-Fi/Macrocell scenario, where the C-plane of  UEs associated with a Wi-Fi access point is provided by the macro eNB in a low licensed frequency band, while the U-plane of UEs associated with a Wi-Fi access point is provided by their serving Wi-Fi access point in a high unlicensed frequency band. The interworking of Wi-Fi and cellular networks benefits from the split of C-plane and U-plane in terms of mobility robustness, service continuity, reduction in cell-planning efforts, energy efficiency, etc.

When considering the backhaul issues of small cells, expensive wired backhaul links may not always be feasible, especially for the dense deployment of small cells. In the meanwhile, in-band wireless backhaul solutions using the licensed spectrum may not be feasible either, because of the scarcity of the licensed spectrum \cite{mmWaveTCOM13}. Recently, the millimeter-wave bands, such as the unlicensed 60 GHz band and the low interference licensed 70 GHz and 80 GHz bands, have been considered as promising candidates for the small cell backhaul solution. This is motivated by the huge frequency bandwidth (globally harmonised over more than 6 GHz millimeter-wave spectrum) that can be exploited, and the spatial isolation supported by highly directional beams. At the same time, the new IEEE 802.11ad standard, a.k.a. WiGig, uses the unlicensed 60 GHz millimeter-wave band to deliver data rates of up to 7 Gbps. This adds a large amount of new frequency bandwidth to existing Wi-Fi products, such as IEEE 802.11n operating in the 2.4 GHz and 5 GHz bands, and IEEE 802.11ac operating in the 5 GHz band.

For providing radio access, the macrocell, small cells and Wi-Fi access points share the unlicensed spectrum. The network architecture in Fig. 1 integrates the coexistence of Wi-Fi and cellular networks and facilitates smart management of data traffic in mobile operators' networks. For instance, the data traffic could be dynamically routed to the optimal radio interface for a particular application and user, with network congestion, reliability, security, and connectivity cost taken into account. As can be seen from Fig. 1, the primary carrier always uses licensed spectrum to transmit control signaling, user data and mobility signalling, while the secondary carrier(s) use unlicensed spectrum to transmit best-effort user data in the downlink and potentially the uplink.

\subsection{Handover Procedure between Wi-Fi and LTE/LTE-A}
\begin{figure}[h]
        \centering
        \includegraphics*[width=15cm]{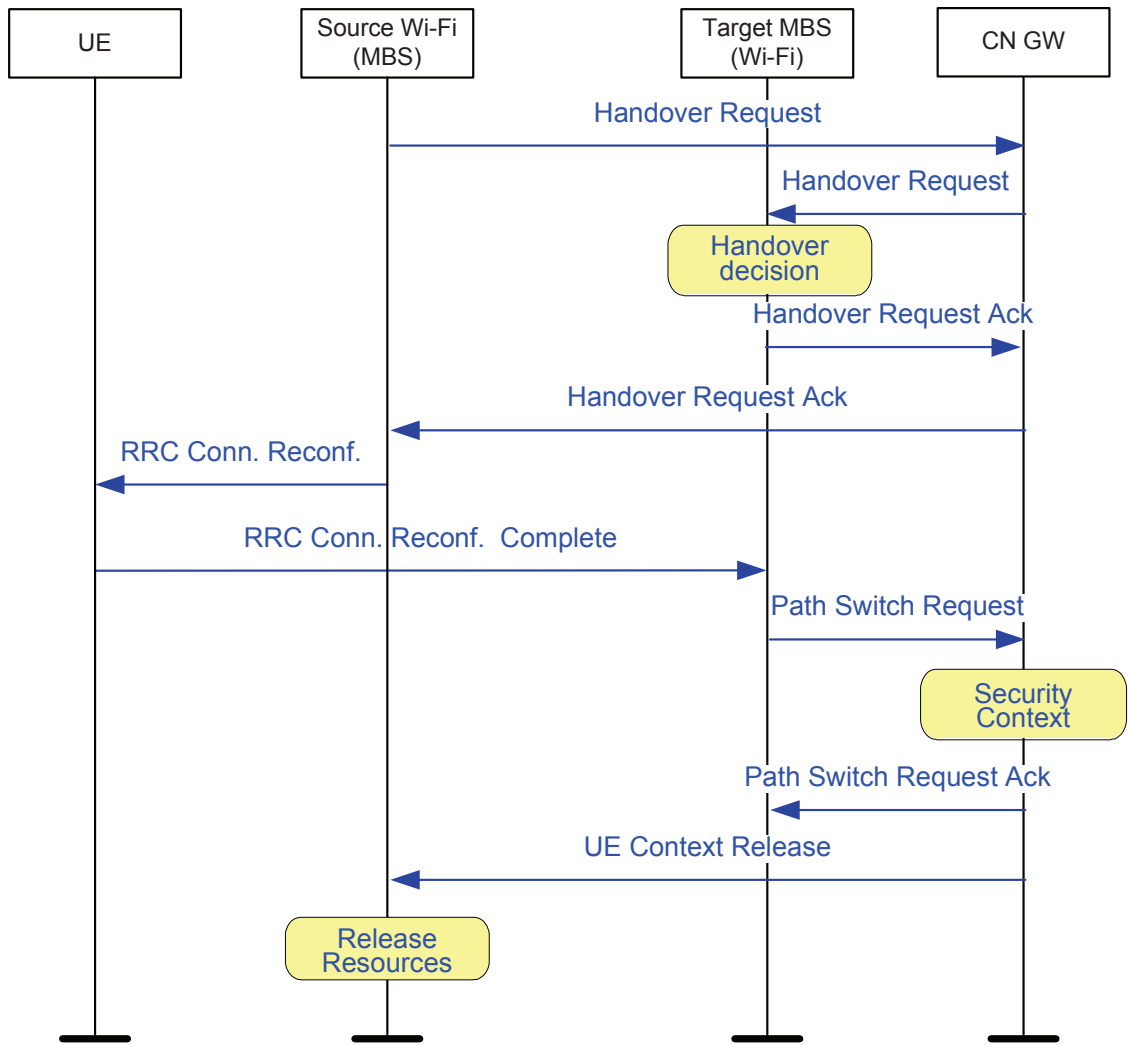}
        \caption{Handover Procedure between Wi-Fi and LTE/LTE-A.}
        \label{HandoverFlow}
\end{figure}

Seamless mobility is one of the key aspects of interworking between Wi-Fi and LTE/LTE-A systems. During the handover process, there should be no package loss or radio link failure in order to ensure the user's QoS. In current Wi-Fi systems, interworking between Wi-Fi and LTE/LTE-A systems is not supported, although it is badly needed due to users' frequent mobility between the coverage areas of Wi-Fi access points and cellular networks. In 3GPP Rel-11, trusted WLAN access to the Enhanced Packet Core (EPC) is based on S2a-based Mobility over GPRS Tunnelling Protocol (SaMOG), which is enhanced in 3GPP Rel-12 to provide traffic steering and mobility between LTE-A and Wi-Fi networks and to optimize the use of network resources. The 3GPP standard TS 23.401 describes seamless and non-seamless handover solutions between 3GPP and non-3GPP access networks. These standards enable users to continue using data services when they pass across macrocells, small cells and Wi-Fi hotspots. In the network architecture shown in Fig. 1, a UE can handover between Wi-Fi and cellular networks through the core network (CN) gateway (GW). In Fig. \ref{HandoverFlow}, the CN GW based handover procedure between a source Wi-Fi (macrocell) and a target macrocell (Wi-Fi) is given. Since the C-plane of Wi-Fi users is managed by the macrocell, the RRC signallings of Wi-Fi users are transmitted from the  macrocell, and the handover signalling overhead between the Wi-Fi access point and the macrocell can be much reduced.

\section{Almost Blank Subframes Allocation}

\subsection{Coexistence without Priority}
In recent years, regulatory bodies are considering the possible coexistence of multiple disparate radio access technologies (RATs) on the same frequency band.  This includes both the licensed bands (e.g., TV spectrum) and unlicensed bands (e.g., amateur spectrum). In countries such as the USA, Canada, and the UK, regulatory efforts are being made to permit the operation of white spaces devices (WSDs). For example, the IEEE 802.22 fixed point-to-point cognitive radio transmissions in TV white spaces, and more recently the IEEE 802.16h wireless broadband protocols.

In the licensed bands, there is a clear notion of the primary and secondary users, whereby spectrum sensing techniques are employed by secondary users to avoid causing interference to primary users.  This can be achieved by identifying primary transmissions using spectrum sensing or geo-location database operations.  However, there is a lack of research activities examining how secondary users associated with different RATs can avoid or mitigate co-channel interference to each other. In the unlicensed bands without the concepts of primary and secondary users, a similar challenge exists between different RATs sharing the same spectrum.

In this article, we refer to the coexistence of two RATs without priority ranking as \emph{coexistence without priority}.
%There is currently discussion regarding this usage of cellular communications (i.e., LTE) in the 2.4GHz unlicensed bands, which has been predominantly used by Wi-Fi communications, sensor networks, and machine-to-machine control signalling.
More specifically, we focus on allowing cellular communications to co-exist with Wi-Fi communications on an equal basis, i.e., no discrimination between primary and secondary users.  %Clearly, the increased mutual interference between the two disparate RATs is a challenge for capacity growth.
What is new here is the coexistence of two disparate RATs that were not designed to be in coexistence, together with the impact of this on the interference map.
Whilst the coexistence of contention based systems have been explored (e.g., IEEE 802.11 and IEEE 802.15 systems) \cite{Yuan09}, the coexistence of a non-contention system (LTE) with a contention system (Wi-Fi) is not well explored, especially when no priority ranking between them is given.  In fact, a reasonable suspicion is that the allocation based transmission protocols of LTE may completely block the collision based protocols of Wi-Fi.  Coupled with the growing density of small cells, this lack of interpretability on the same spectrum band can cause severe capacity issues.

%Detection of coexistence is not discussed in great detail here, except to say that
In multiple RAT coexistence, communication protocols can operate in either their default normal mode or a coexistence mode. The latter is triggered when another RAT is sensed nearby and action is needed. Coexistence mechanisms can be divided into two groups: i) those that require message exchange between nodes or RATs, and ii) those that do not. In general, cross-RAT coordination is difficult due to the disparate protocol development processes and vendor differences. Therefore, in the following sub-section we will review a non-collaborative coexistence mechanism that allows LTE to co-exist with Wi-Fi in the unlicensed spectrum.

\subsection{Random Almost Blank Subframe Allocation}

In \cite{Almeida13ICC}, autonomous (without coordination) coexistence between LTE and Wi-Fi was achieved by LTE transmitting ABSs under a 3GPP Rel-10 time-division-duplex (TDD) scheme. The ABSs are subframes with reduced power or content. They are backwards compatible with 3GPP Rel-8 and Rel-9 in that several synchronization channels remain (e.g., common reference signals).  For interference avoidance between cells of the same RAT, ABSs are triggered by coordination messages between eNBs via the X2 interface.  The frequency of ABS transmissions can be adapted to the time-varying interference environment.
For interference avoidance between cells of different RATs, coordination messaging between cells of different RATs is challenging for the previously mentioned reasons. Therefore, ABSs are transmitted randomly at some rate without coordination and without the need for backwards compatibility with previous releases on the unlicensed bands \cite{Almeida13ICC}.
The central conceit to this idea is that during the random ABSs, the Wi-Fi access points can detect the channel vacancy and transmit following its contention based protocol. Accordingly, the allocation nature of LTE transmissions can be suppressed in a way that avoids coordination or spectrum sensing, but at the cost of decreased LTE spectral efficiency and network capacity.
\begin{figure}[t]
    \centering
    \includegraphics[width=1.00\linewidth]{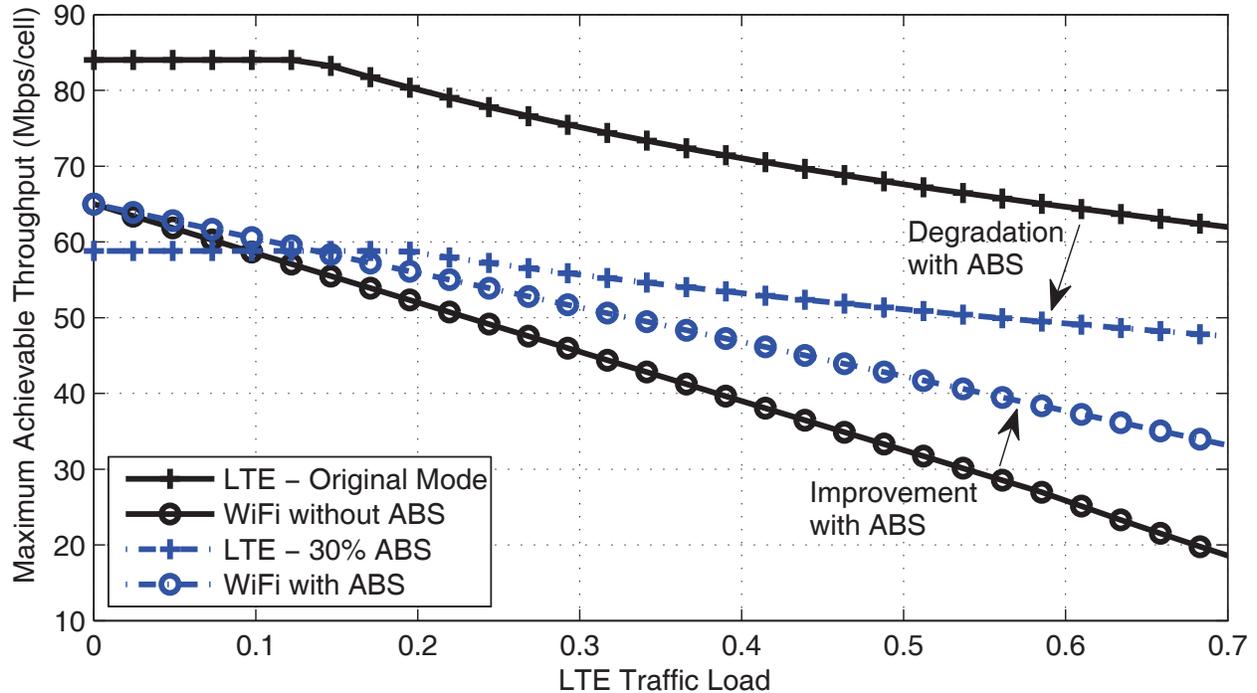}
    \caption{Maximum achievable throughput (Mbps per cell) for an LTE or Wi-Fi cell under different normalized traffic loads (normalized to cell capacity), with LTE operating with: a) original mode, and b) ABS.}
    \label{ABS}
\end{figure}
\begin{table}
    \caption{Simulation Parameters.}
    \begin{center}
        \begin{tabular}{|l|l|}
          \hline
          \emph{Parameter}              & \emph{Value}              \\
          \hline
          LTE Carrier Frequency         & 2100 MHz                  \\
          Wi-Fi Carrier Frequency       & 2400 MHz                  \\
          LTE Cell Density              & 3 per km$^{2}$            \\
          Wi-Fi Density                 & 300 per km$^{2}$          \\
          LTE Cell Transmit Power       & 40 W                      \\
          Wi-Fi Transmit Power          & 1 W                       \\
          Pathloss                      & 3GPP Urban Micro          \\
          Peak LTE Throughput           & 84 Mbits/s (64 QAM SISO)  \\
          Peak IEEE 802.11n Throughput       & 65 Mbits/s (64 QAM SISO)  \\
          \hline
        \end{tabular}
    \end{center}
    \label{Parameters}
\end{table}

In Fig. \ref{ABS}, the maximum achievable throughput for an LTE or Wi-Fi (IEEE 802.11n) cell under different normalized traffic loads are plotted, with LTE operating either under the original mode or with 30\% of the subframes randomly selected as ABSs.  The parameters used in the simulation can be found in Table~\ref{Parameters}.  The results show that under the original mode, LTE cell capacity saturates at around 84 Mbps due to discrete modulation and coding schemes employed (64QAM with Turbo coding), while Wi-Fi cell capacity saturates at 64 Mbps. As the LTE traffic load increases, the capacity of all cells falls due to increased radio resource usage and the resulting increased interference. LTE cell capacity decays slower than Wi-Fi capacity. The \emph{super-linear} degradation of Wi-Fi capacity may finally lead to 0 Mbps.  By employing 30\% ABSs in LTE, the Wi-Fi capacity is improved significantly at high LTE traffic loads, while the LTE cell capacity falls by 10--24 Mbps. Note that the capacity degradation rates for both LTE and Wi-Fi become slower with LTE ABS transmissions, because random ABSs mitigate both cross-tier and co-tier inter-cell interference. It is worth noting that the overall aggregate capacity of LTE and Wi-Fi is actually reduced with ABS, indicating that the random ABS mechanism benefits fairness instead of overall capacity.

In summary, ABSs can be transmitted randomly by LTE transmitters to allow the spectrum-sharing coexistence of allocation-based LTE transmissions and contention-based Wi-Fi transmissions
with an improved fairness between them, but at the cost of decreased overall aggregate capacity of LTE and Wi-Fi networks.
%The ABS mechanism actually decreases the overall combined network capacity, but it does so to allow a greater fairness ratio between LTE and Wi-Fi.

\section{Interference Avoidance with Neighborhood Wi-Fi Density Estimation}

\subsection{Inference Framework}

It has been shown that avoiding co-channel interference in a network with a high interference intensity can improve the long-term system throughput~\cite{Liang12}. However, coordinating interference avoidance on the radio resource management (RRM) level typically requires a large volume of coordination information exchanged between multiple base stations (BSs) via the X2 interface. Specifically, each BS in an OFDMA system needs to know whether its neighboring BSs are transmitting on each available radio resource block. This level of coordination taxes the \emph{backhaul capacity}, while any \emph{delay} in information sharing may cause the interference avoidance performance to falter.

In \cite{Guo14ICT}, an interference estimation technique that does not require information sharing between BSs or UEs was devised based on each BS sensing the spectrum and estimating the number of co-channel transmissions in a defined observation zone. By estimating the number of co-channel transmitters and knowing the cell density in the region, the average channel quality at \emph{any random point} in a coverage area can be inferred. As the expressions are tractable, the computational complexity is extremely low. The methodology can be applied to a $\mathsf{K}$-tier heterogeneous network by leveraging a stochastic geometry framework and an opportunistic interference reduction scheme, which was shown to approach the interference estimation accuracy achievable by information exchange on the X2 interface \cite{Guo14ICT}.
\begin{figure}[t]
    \centering
    \includegraphics[width=0.95\linewidth]{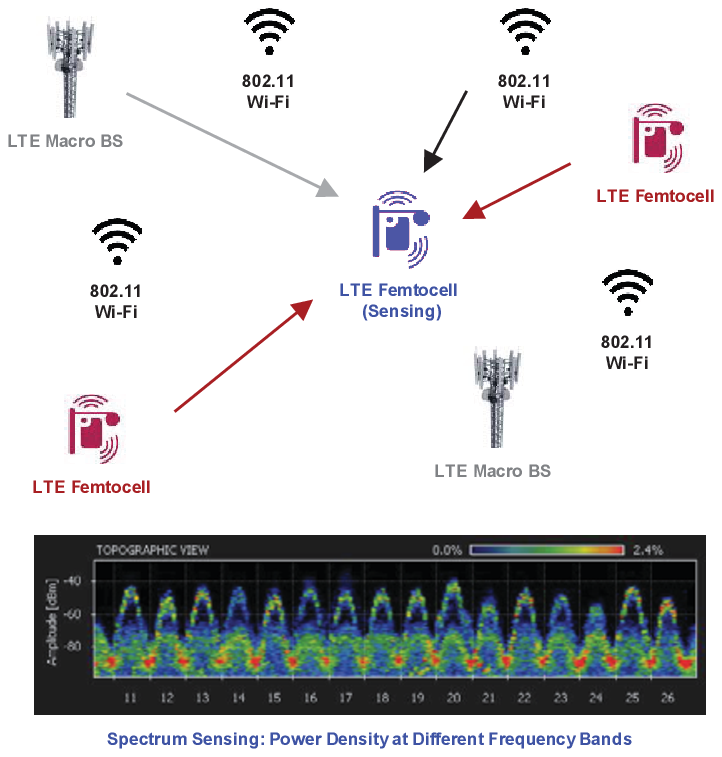}
    \caption{Illustration of a femtocell inferring the number of co-channel transmitters from a 3-tier cellular and Wi-Fi heterogeneous network by sensing the received power spectrum \cite{Guo14ICT}.}
    \label{Zone}
\end{figure}

The inference framework assumes that each cell is equipped with a spectrum sensing device \footnote{Low cost spectrum sensing equipment for 2--5GHz is now readily available}. On each frequency band $f$, the sensor at each cell (located at distance $h$ from the BS) is able to detect the power density $P_{f}$ from all co-channel transmitters in an unbounded region.  Given knowledge of the spatial distribution of co-channel cell deployments \cite{Haenggi05}, the density of co-channel transmissions $\lambda_{f}$ can be inferred from the $P_{f}$ measurements \cite{Guo14ICT}:
\begin{equation} \begin{split}\label{Inference1}
\lambda_{f} \propto \frac{\sqrt{P_{f}/P}}{Q(h,\alpha)}
\end{split}\end{equation} where $\alpha$ is the pathloss distance exponent, $P$ is the average transmit power of the BSs, and the function $Q(h,\alpha)$ is given in \cite{Guo14ICT}.  Without loss of generality, this inference framework can be applied to a $\mathsf{K}$-tier heterogeneous network comprised of macrocells, femtocells and Wi-Fi access points.  Fig. \ref{Zone} illustrates how a femtocell infers the number of co-channel transmitters in a 3-tier heterogeneous network by sensing the received power spectrum.  In this illustrative example, the estimated transmitter activities on the considered frequency band are: 50\% of LTE macrocells, 100\% of LTE femtocells, and 25\% of Wi-Fi access points. This spectrum sensing mechanism is not able to know which cells are transmitting, but it provides a statistical notion for a BS to infer the channel quality of a served user.

Based on the inferred density of co-channel transmissions $\lambda_{f}$ in the vicinity, the signal-to-interference ratio (SIR) on frequency band $f$ at distance $d$ away from the sensing BS is estimated as:
\begin{equation} \begin{split}\label{Inference2}
\text{SIR}_{f,d} \propto P_{f}^{-1} \bigg[ \frac{Q(h,\alpha)}{Q(d,\alpha)} \bigg]^{2},
\end{split}\end{equation} where the constant of proportionality is the received signal strength.

\subsection{Simulation Results}

\begin{figure}[t]
    \centering
    \includegraphics[width=16cm]{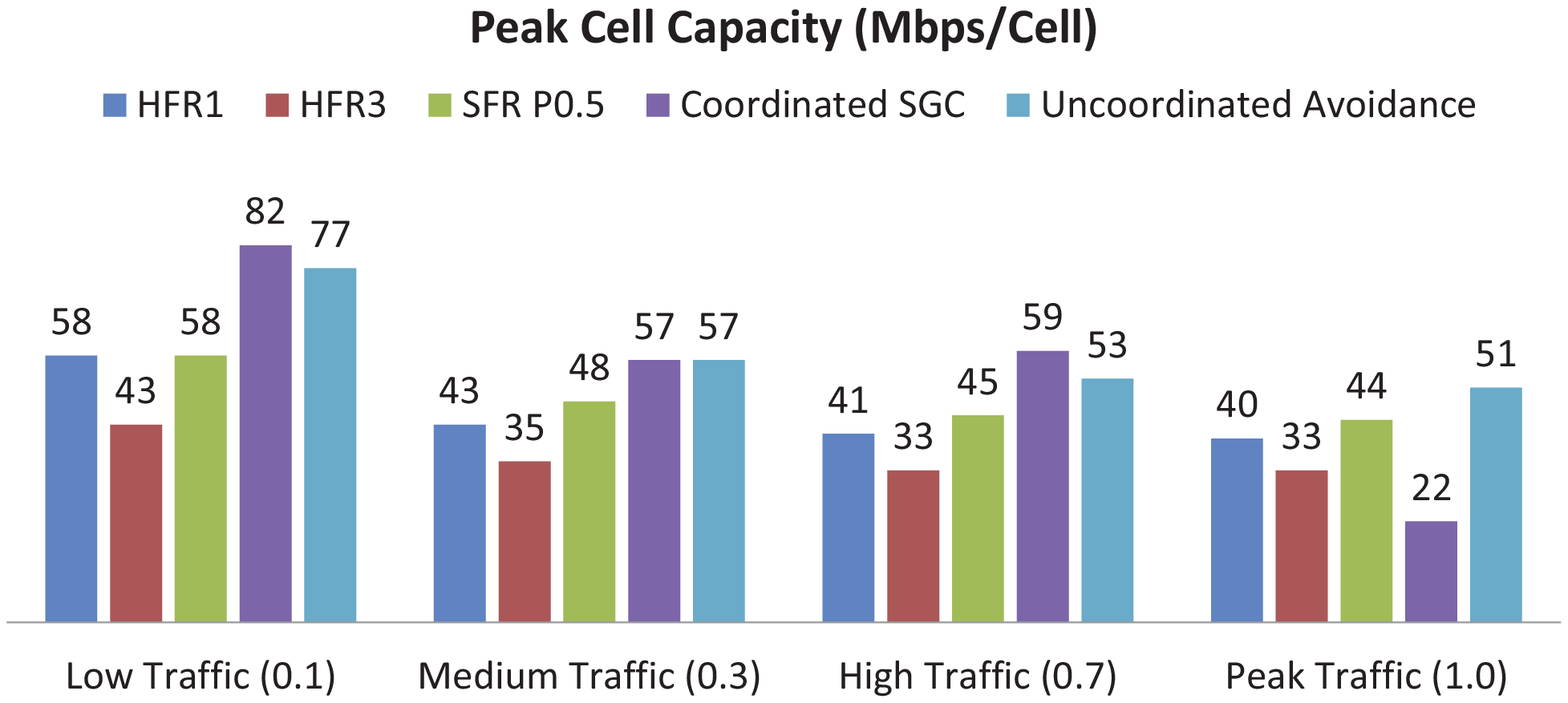}
    \caption{Plot of peak cell capacity versus different normalized cell traffic loads for a variety of static and dynamic interference mitigation schemes \cite{Guo14ICT, Turyagyenda12IET}.}
    \label{Results}
\end{figure}

Fig. \ref{Results} shows the simulation results of peak cell capacity versus normalized cell traffic load for a variety of static and dynamic interference mitigation schemes \cite{Guo14ICT, Turyagyenda12IET}. The baseline is a hard frequency reuse 1 (HFR1) scheme, which shows a peak capacity of 58 Mbps/cell when the cells are unloaded (i.e., minimum inter-cell interference).  This value falls steadily to 40 Mbps/cell for fully loaded cells without interference mitigation (i.e., maximum inter-cell interference).  A similar trend exists for HFR3. The soft frequency reuse with a power backoff factor 0.5 (SFR P0.5) performs better than the previous two schemes.  We can see from Fig. \ref{Results} that the TDD-based sequential game coordinated (SGC) interference avoidance scheme achieves a much higher peak cell capacity than the HFR and SFR schemes at low and medium cell traffic loads. The uncoordinated interference avoidance scheme proposed in \cite{Guo14ICT} provides the highest peak cell capacity at high traffic loads and achieves over 90\% the peak cell capacity of the SGC interference avoidance scheme that requires channel state information.

\subsection{Discussion and Challenges}

The disadvantage with a TDD-based coordinated interference avoidance scheme is the need for two prerequisites \cite{Turyagyenda12IET}: (1) cell pairing or clustering; (2) static or dynamic assignment of cell priority.  Effective cell pairing often involves the association of cells that are dominant interferers to each other. However, this may not always be the case. For example, the antenna bore-sight of BS $A$ is pointing at BS $B$, but the antenna bore-sight of BS $B$ is pointing at a direction away from BS $A$. Alternatively, two cells that are closest to each other will be paired together.  Cell priority assignment refers to the process of assigning different transmission priorities to cells.  Random access, traffic weighted, and QoS weighted cell priority assignments have been considered in the literature.

\section{Conclusion}

In this article, we have looked into the potentials and challenges associated with coexisting Wi-Fi systems and heterogeneous cellular networks sharing the unlicensed spectrum. We have introduced the network architecture for LTE/LTE-A small cells to exploit the unlicensed spectrum already used by Wi-Fi systems. The ABS mechanism and an interference avoidance scheme have been presented to mitigate the interference between Wi-Fi and LTE/LTE-A systems when both transmitting in the same unlicensed spectrum. Simulation results have shown that with the proper use of ABS mechanism and interference avoidance schemes, heterogeneous and small cell networks can improve their capacity by using the unlicensed spectrum used by Wi-Fi systems without affecting the performance of Wi-Fi.

\section*{Acknowledgment}
This work was supported by the National Natural Science Foundation of China (61471025), the Fundamental Research Funds for the Central Universities (Grant No. ZY1426), and the Interdisciplinary Research Project in BUCT.

\begin{IEEEbiography}{Haijun Zhang} (M'13) received his Ph.D. degree from Beijing University of Posts Telecommunications (BUPT). He is currently a Postdoctoral Research Fellow in Department of Electrical and Computer Engineering, the University of British Columbia (UBC). He was an Associate Professor in College of Information Science and Technology, Beijing University of Chemical Technology. From 2011 to 2012, he visited Centre for Telecommunications Research, King's College London, London, UK, as a joint PhD student and Research Associate.  He has published more than 50 papers and has authored 2 books. He serves as the editors of Journal of Network and Computer Applications, Wireless Networks (Springer), and KSII Transactions on Internet and Information Systems. He served as Symposium Chair of GAMENETS'2014 and Track Chair of ScalCom'2015. He also serves or served as TPC members of many IEEE conferences, such as Globecom and ICC. His current research interests include 5G, Resource Allocation, Heterogeneous Small Cell Networks and Ultra-Dense Networks.
\end{IEEEbiography}

\begin{IEEEbiography}{Xiaoli Chu}(M'05) is a Lecturer in the Department of Electronic and Electrical Engineering at the University of Sheffield, UK. She received the B.Eng. degree in Electronic and Information Engineering from Xi'an Jiao Tong University in 2001 and the Ph.D. degree in Electrical and Electronic Engineering from the Hong Kong University of Science and Technology in 2005. From Sep. 2005 to Apr. 2012, she was with the Centre for Telecommunications Research at King¡¯s College London. She is the leading editor/author of the book Heterogeneous Cellular Networks ¨C Theory, Simulation and Deployment, Cambridge University Press, May 2013. She is Guest Editor of the Special Section on Green Mobile Multimedia Communications for IEEE Transactions on Vehicular Technology (Jun. 2014) and the Special Issue on Cooperative Femtocell Networks for ACM/Springer Journal of Mobile Networks \& Applications (Oct. 2012). She is Co-Chair of Wireless Communications Symposium for the IEEE International Conference on Communications 2015 (ICC'15), was Workshop Co-Chair for the IEEE International Conference on Green Computing and Communications 2013 (GreenCom'13), and has been Technical Program Committee Co-Chair of several workshops on heterogeneous networks for IEEE GLOBECOM, WCNC, PIMRC, etc.
\end{IEEEbiography}

\begin{IEEEbiography}
{Weisi Guo} (M'11) received his M.Eng., M.A. and Ph.D. degrees from the University of Cambridge. He is currently an Assistant Professor and Co-Director of Cities research theme at the School of Engineering, University of Warwick.  In recent years, he has published over 50 IEEE papers and won a number of awards and grants.  He is the author of the VCEsim LTE System Simulator, and his research interests are in the areas of heterogeneous networks, self-organization, energy-efficiency, and molecular communications.
\end{IEEEbiography}

\begin{IEEEbiography}
{Siyi Wang} (M'13) received his Ph.D degree in wireless communications from the University of Sheffield, UK in 2014. He was a research member of the project ``Core 5 Green Radio'' funded by Virtual Centre of Excellence (VCE) and UK EPSRC. He is currently a lecturer at Xi'an Jiaotong-Liverpool University (XJTLU). He has led a team and won the second prize of the IEEE Communications Society Student Competition ``Communications Technology Changing the World". He has published over 25 IEEE papers in the past 3 years in the area of 4G cellular networks, molecular communications and mobile sensing,  and won an IEEE best conference paper award. His research interests include: molecular communications, indoor-outdoor network interaction, small cell deployment, device-to-device (D2D) communications, machine learning, stochastic geometry, theoretical frameworks for complex networks and urban informatics.
\end{IEEEbiography}

\end{document}